# 4D Pyritohedral Symmetry with Quaternions, Related Polytopes and Lattices

**Mehmet Koca[a], Nazife Ozdes Koca[a*] and Amal Al-Qanobi[a]**

[a] Department of Physics, College of Science, Sultan Qaboos University, P.O. Box 36, Al-Khoud, 123 Muscat, Sultanate of Oman. *Correspondence e-mail: nazife@squ.edu.om

**Abstract**

We describe extension of the pyritohedral symmetry to 4-dimensional Euclidean space and present the group elements in terms of quaternions. It turns out that it is a maximal subgroup of both the rank-4 Coxeter groups $W(F_4)$ and $W(H_4)$ implying that it is a group relevant to the crystallography as well as quasicrystallographic structures in 4-dimensions. First we review the pyritohedral symmetry in 3 dimensional Euclidean space which is a maximal subgroup both in the Coxeter-Weyl groups $W(B_3) \approx Aut(D_3)$ and $W(H_3)$. The related polyhedra in 3-dimensions are the two dual polyhedra pseudoicosahedron-pyritohedron and the pseudo icosidodecahedron. In quaternionic representations it finds a natural extension to the 4-dimensions. The related polytopes turn out to be the pseudo snub 24-cell and its dual polytope expressed in terms of a parameter $x$ leading to snub 24-cell and its dual in the limit where the parameter $x$ takes the golden ratio. It turns out that the relevant lattice is the root lattice of $W(D_4)$

## 1. Introduction

Lattices in higher dimensions described by the affine Coxeter groups, when projected into lower dimensions, may represent the quasicrystal structures (Katz and Duneau, 1986; Elser, 1985; Baake et al., 1990a; Baake et al., 1990b, Koca et al. 2014a, Koca et al. 2015a). Projection of the 4-dimensional self-dual lattice $F_4$ in a 2 dimensional subspace can describe a 12-fold symmetric aperiodic tiling of the plane (Koca et al. 2014b). It was already known that the $A_4$ lattice projects into the aperiodic lattice with 5-fold symmetry (Baake et al., 1990a). There is no doubt that the projections of the higher dimensional lattices may have some physical implications. Let us remember that the quasicrystallographic Coxeter group $W(H_4)$ can be obtained as the subgroup of the Coxeter-Weyl group of $W(E_8)$ (Koca et. al., 2001) describing the densest sphere packing lattice in 8-dimensions (Conway and Sloane, 1988). $E_8$ as a Lie group may represent the unified symmetry of the fundamental interaction besides gravity (Bars and Gunaydin, 1980; Koca, 1981) or it may represent all fundamental interactions in the context of heterotic superstring theory. The exceptional Coxeter-Weyl group $W(F_4)$ describes the symmetry of the unique self-dual polytope, the 24-cell, which is the Voronoi cell (Wigner-Seitz cell) of the $F_4$ lattice. The noncrystallographic Coxeter group $W(H_4)$ is the symmetry of the famous 600-cell and its dual 120-cell (Coxeter, 1973; Koca et al. 2007a). In what follows we will construct the 4D pyritohedral group from the $D_4$ diagram; in technical terms it is the group $\left(W(D_4)/C_2\right):S_3$ (Koca et al. 2012) of order 576 which can be expressed in terms of quaternions and we will determine its orbits as the pseudo snub 24-cell and its dual polytope.

Although it was so fascinating, the quaternionic representations of the symmetries of the 4D and 3D Euclidean spaces have not been studied for a long time. The finite subgroups of quaternions lead to the classifications of the finite subgroups of $O(4)$ in terms of pairs of quaternionic finite subgroups (du Val, 1964; Conway & Smith, 2003). It turned out that all rank-4 Coxeter-Weyl groups can be represented in compact forms generated by quaternion pairs (Koca et al., 2010).

This paper is organized as follows. In Section 2 we review the quaternionic representations of the $O(4)$ and $O(3)$ symmetries and point out some important finite subgroups. The quaternionic representations of the Coxeter groups $W(F_4)$ and $W(H_4)$ are briefly reviewed. Section 3 deals with the quaternionic description of the symmetries generated by Coxeter-Dynkin diagrams $D_3$ and $B_3$ in which the simple roots (lattice generating vectors) expressed in terms of imaginary quaternionic units. The quaternionic representation of the pyritohedral group obtained from the $D_3$ diagram and its orbits describing the vertices of the pseudoicosahedron are studied in an earlier paper (Koca et al., 2015b). In Section 4 we construct the group $\left(W(D_4)/C_2\right):S_3$ and apply it to a vector expressed in terms of a free parameter to generate the vertices of a pseudo snub 24-cell. The cell structure (facets) of the pseudo snub 24-cell has been worked out and the vertices of the dual polytope is constructed. We construct also the pseudo snub 24-cell based on a different quaternionic representation of the $D_4$ diagram leading to another construction of the group $\left(W(D_4)/C_2\right):S_3$ in terms of quaternions. The Fibonacci chain of the pseudo snub 24-cells leading to the snub 24-cell is introduced. Finally in Section 5 we present a brief discussion on the physical implications of our technique when the lattice structure is projected into either 3-dimensions or 2-dimensions.

## 2. Quaternions and orthogonal transformations in 3 and 4 dimensions

A quaternion is a hypercomplex number with three imaginary units $i, j, k$ satisfying the relation $i^2 = j^2 = k^2 = ijk = -1$ (Hamilton, 1843). Hereafter we shall use a different notation for the imaginary units by redefining them as $e_1 = i, e_2 = j, e_3 = k$. A real unit quaternion $q = q_0 + q_i e_i$, $(i = 1,2,3)$ and its quaternion conjugate $\bar{q} = q_0 - q_i e_i$ satisfy the norm equation $q\bar{q} = \bar{q}q = q_0^2 + q_1^2 + q_2^2 + q_3^2 = 1$. In terms of new imaginary units the Hamilton's relation can be written in a compact form

$$e_i e_j = -\delta_{ij} + \varepsilon_{ijk} e_k, \quad (i, j, k = 1,2,3) \tag{1}$$

where $\delta_{ij}$ and $\varepsilon_{ijk}$ represent respectively the Kronecker and Levi-Civita symbols and summation over the repeated indices is implicit. Scalar product of two quaternionic vectors can be defined as

$$(p, q) = \frac{1}{2}(\bar{p}q + \bar{q}p) = \frac{1}{2}(p\bar{q} + q\bar{p}). \tag{2}$$

With this scalar product quaternions generate the four-dimensional euclidean space.

Let $p$ and $q$ be two arbitrary unit quaternions $p\bar{p} = q\bar{q} = 1$ with six real parameters. The transformations of an arbitrary quaternion in the manner $t \to ptq$ and $t \to p\bar{t}q$ define an orthogonal transformation of the group $O(4)$. It is clear that the above transformations preserve the norm $t\bar{t} = t_0^2 + t_1^2 + t_2^2 + t_3^2$ which can be represented by the abstract group operations in the form of notations

$$t \to ptq := [p, q]t, \quad t \to p\bar{t}q := [p, q]^* t. \tag{3}$$

One can also drop *t*, then, the pairs of quaternions define a set closed under multiplication (Koca et al, 2001). The inverse elements take the forms

$$[p,q]^{-1} = [\bar{p},\bar{q}], \quad ([p,q]^*)^{-1} = [\bar{q},\bar{p}]^*. \qquad (4)$$

With the choice of $q = \bar{p}$ the orthogonal transformations define a three parameter subgroup $O(3)$. This restriction allows the definition of a quaternion consisting of a scalar component $Sc(t) = t_0$ and a vector component $\text{Vec}(t) = t_1 e_1 + t_2 e_2 + t_3 e_3$. The transformation $[p,\bar{p}]$ and $[p,\bar{p}]^*$ leaves the $Sc(t) = t_0$ invariant. Therefore in 3D Euclidean space one can assume, without loss of generality that the quaternions consist of only vector components that is $t = t_1 e_1 + t_2 e_2 + t_3 e_3$ satisfying $\bar{t} = -t$. With this restriction to the 3D space the element $[p,\bar{p}]^*$ takes a simple form $[p,\bar{p}]^* \Rightarrow [p,-\bar{p}]$. Therefore the transformations of the group $O(3)$ can be written as $[p,\pm\bar{p}]$ where the group element $[p,\bar{p}]$ represents the rotations around the vector $\text{Vec}(p) = (p_1, p_2, p_3)$ and $[p,-\bar{p}]$ is a rotary inversion (Coxeter, 1973).

In the references (Conway and Smith, 2003 and du Val, 1964) the classifications of the finite subgroups of quaternions and the related finite subgroups of the group of orthogonal transformations $O(3)$ are listed. The relevant finite subgroups of quaternions in our work are the binary tetrahedral group and the binary octahedral group. The binary tetrahedral group $T$ is given by the set of quaternions

$$T = \{\pm 1, \pm e_1, \pm e_2, \pm e_3, \frac{1}{2}(\pm 1 \pm e_1 \pm e_2 \pm e_3)\} \qquad (5)$$

with 24 elements. As we will point out later that the same set represents the vertices of the 24-cell, a self-dual polytope in 4D [Coxeter, 1973]. The 24-cell represented by another set of 24 quaternions are given by

$$T' = \{\frac{1}{\sqrt{2}}(\pm 1 \pm e_1), \frac{1}{\sqrt{2}}(\pm e_2 \pm e_3), \frac{1}{\sqrt{2}}(\pm 1 \pm e_2), \frac{1}{\sqrt{2}}(\pm e_3 \pm e_1), \frac{1}{\sqrt{2}}(\pm 1 \pm e_3), \frac{1}{\sqrt{2}}(\pm e_1 \pm e_2)\} \quad (6)$$

which does not form a group by itself. Since these two sets satisfy the relations $T'T' \subset T$ and $T'T \subset T'$ their union $O = T \cup T'$ forms another finite subgroup of quaternions called the binary octahedral group. In an earlier publication (Koca et.al., 2001) we have proved that the Coxeter-Weyl group $W(F_4)$ of order 1152 can be represented as

$$W(F_4) = \{[T,T] \cup [T,T]^* \cup [T',T'] \cup [T',T']^*\}. \qquad (7)$$

The automorphism group of $F_4$ is given by the set

$$Aut(F_4) = \{[O,O] \cup [O,O]^*\}. \qquad (8)$$

Let us clarify our notation: we mean for example by $[O,O]$ an arbitrary pair of unit quaternions $[p,q]$ where $p,q \in O$. The set of 120 quaternions $I = T \cup S$ (for an explicit form of S see Koca et al. 2007a) where $S$ represents 96 quaternions which is invariant under the group transformation $TST \in S$ and $T\bar{S}T \in S$. The quasicrystallographic Coxeter group $W(H_4) = \{[I,I] \cup [I,I]^*\}$ can be further decomposed as

$$W(H_4) = \{[T,T] \cup [T,S] \cup [S,T] \cup [S,S] \cup \\ [T,T]^* \cup [T,S]^* \cup [S,T]^* \cup [S,S]^*\}. \tag{9}$$

We note that the group $\{[T,T] \cup [T,T]^*\}$ of order 576 is a maximal subgroup in the both groups $W(F_4)$ and $W(H_4)$. In Section 4 we will generate this group from the $D_4$ diagram and study its orbits.

## 3. The Coxeter-Weyl groups $W(D_3)$ and $W(B_3)$ represented by quaternions

The Coxeter-Weyl groups are generated by reflections with respect to some hyperplanes represented by the vectors (also called roots in the literature of Lie Algebra). Let us denote by $\alpha_1, \alpha_2, ..., \alpha_n$ the simple roots of the Coxeter-Dynkin diagram. Denote by $r_{\alpha_i} := r_i$, $(i = 1,2,...,n)$ the reflection operator with respect to the hyperplane orthogonal to the simple root $\alpha_i$. Then the presentation of the Coxeter-Weyl group $W(G)$ is given by

$$W(G) = \langle r_1, r_2, ..., r_n | (r_i r_j)^{m_{ij}} = 1 \rangle \tag{10}$$

where $m_{ij}$ is an integer with $m_{ii} = 1$, $m_{ij} = 2,3,4$ and 6 for $i \neq j$ representing respectively no line, one line, two lines (or label 4) and three lines (or label 6) between the nodes of the Coxeter-Dynkin diagrams which determine the crystallographic point groups in an arbitrary euclidean space. There are also quasicrystallographic Coxeter groups $I_2(n)$, $(n \geq 5, n \neq 6)$, $H_3$ and $H_4$. The list of Coxeter-Weyl groups can be found in many fundamental references (Coxeter and Moser, 1965; Bourbaki, 1968; Humphreys, 1990).

The reflection of an arbitrary vector $\lambda$ with respect to the hyperplane represented by the simple root $\alpha_i$ is given by

$$r_i \lambda = \lambda - \frac{2(\lambda, \alpha_i)}{(\alpha_i, \alpha_i)} \alpha_i, \quad (i = 1, 2, ..., n). \tag{11}$$

The Cartan matrix $C$ (Gram matrix in the lattice terminology) with the matrix elements $C_{ij} = \frac{2(\alpha_i, \alpha_j)}{(\alpha_j, \alpha_j)}$ and the metric $G$ defined by the matrix elements $G_{ij} = (C^{-1})_{ij} \frac{(\alpha_j, \alpha_j)}{2}$ are important for the description of the Coxeter-Weyl groups and the corresponding lattices. The matrices $C$ and $G$ represent the Gram matrices of the direct lattice and the reciprocal lattice respectively (Conway and Sloane, 1988). We take the roots $\alpha_i$ as the generating vectors of the direct lattice. The weights $\omega_i$ spanning the dual space and satisfying the scalar product $(\omega_i, \omega_j) = G_{ij}$ and $(\omega_i, \breve{\alpha}_j) = \delta_{ij}$ with $\breve{\alpha}_j := \dfrac{2\alpha_j}{(\alpha_j, \alpha_j)}$ correspond to the generating vectors of the reciprocal lattice. Now we briefly review the cubic lattices associated with the Coxeter-Weyl groups $W(D_3)$ and $W(B_3)$.

## 3.1. Construction of the fcc and the bcc lattices as the affine Coxeter-Weyl groups $W_a(D_3)$

The Coxeter-Dynkin diagram of $D_3$ with the quaternionic simple roots is given by Fig. 1.

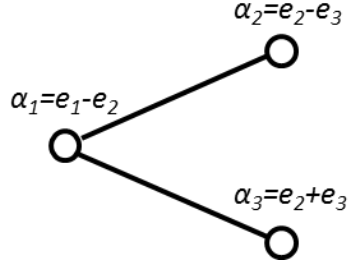

**Figure 1.** The Coxeter-Dynkin diagram $D_3$ with quaternionic simple roots. The angle between two connected roots is $120^0$ otherwise they are orthogonal.

An arbitrary quaternion $\lambda$ when reflected by the operator $r_\alpha$ with respect to the hyperplane orthogonal to the quaternion $\alpha$ the formula (11) can be written in terms of quaternions as

$$r_\alpha \lambda = -\dfrac{\alpha}{\sqrt{2}} \bar{\lambda} \dfrac{\alpha}{\sqrt{2}} := [\dfrac{\alpha}{\sqrt{2}}, -\dfrac{\alpha}{\sqrt{2}}]^* \lambda. \qquad (12)$$

The Cartan matrix of the Coxeter-Dynkin diagram $D_3$ and its inverse matrix are given respectively by the matrices

$$C = \begin{bmatrix} 2 & -1 & -1 \\ -1 & 2 & 0 \\ -1 & 0 & 2 \end{bmatrix}, \quad C^{-1} = \dfrac{1}{4}\begin{bmatrix} 4 & 2 & 2 \\ 2 & 3 & 1 \\ 2 & 1 & 3 \end{bmatrix}. \qquad (13)$$

The generators of the Coxeter group $W(D_3)$ are then given in the notation of (12) by

$$r_1 = [\frac{1}{\sqrt{2}}(e_1-e_2), -\frac{1}{\sqrt{2}}(e_1-e_2)]^* = [\frac{1}{\sqrt{2}}(e_1-e_2), \frac{1}{\sqrt{2}}(e_1-e_2)],$$

$$r_2 = [\frac{1}{\sqrt{2}}(e_2-e_3), -\frac{1}{\sqrt{2}}(e_2-e_3)]^* = [\frac{1}{\sqrt{2}}(e_2-e_3), \frac{1}{\sqrt{2}}(e_2-e_3)], \quad (14)$$

$$r_3 = [\frac{1}{\sqrt{2}}(e_2+e_3), -\frac{1}{\sqrt{2}}(e_2+e_3)]^* = [\frac{1}{\sqrt{2}}(e_2+e_3), \frac{1}{\sqrt{2}}(e_2+e_3)].$$

It is straightforward to check that the reflection generators transform the quaternionic imaginary units as follows:

$$r_1 : e_1 \leftrightarrow e_2, e_3 \to e_3; \quad r_2 : e_1 \to e_1, e_2 \leftrightarrow e_3; \quad r_3 : e_1 \to e_1, e_2 \leftrightarrow -e_3. \quad (15)$$

The Coxeter-Weyl group $W(D_3)$ of order 24 isomorphic to the tetrahedral group takes a simple form in terms of quaternions

$$W(A_3) = \{[p, \bar{p}] \cup [t, -\bar{t}]\}, \quad p \in T, \, t \in T', \quad (16)$$

or more compactly by the notation $W(A_3) = \{[T, \bar{T}] \cup [T', -\bar{T}']\}$. This hybrid group is also called the tetra-octahedral group (Conway and Smith, 2003). Here $T$ and $T'$ are the sets of quaternions given in (5-6).

The choice of simple roots by $\alpha_1 = e_1 - e_2, \alpha_2 = e_2 - e_3, \alpha_3 = e_2 + e_3$ in Fig. 1 determines the weight vectors as

$$\omega_1 \equiv (100) = e_1, \quad \omega_2 \equiv (010) = \frac{1}{2}(e_1+e_2-e_3), \quad \omega_3 \equiv (001) = \frac{1}{2}(e_1+e_2+e_3). \quad (17)$$

Using the orbit definition $W(D_3)(a_1 a_2 a_3) := (a_1 a_2 a_3)_{D_3}$ some of the important orbits can be found:

$$(100)_{D_3} = \{\pm e_1, \pm e_2, \pm e_3\},$$

$$(010)_{D_3} = \{\frac{1}{2}(-e_1-e_2-e_3), \frac{1}{2}(-e_1+e_2+e_3), \frac{1}{2}(e_1+e_2-e_3), \frac{1}{2}(e_1-e_2+e_3)\},$$

$$(001)_{D_3} = \{\frac{1}{2}(e_1+e_2+e_3), \frac{1}{2}(e_1-e_2-e_3), \frac{1}{2}(-e_1-e_2+e_3), \frac{1}{2}(-e_1+e_2-e_3)\}, \quad (18)$$

$$(011)_{D_3} = \{\pm e_1 \pm e_2, \pm e_2 \pm e_3, \pm e_3 \pm e_1\}.$$

The orbits respectively represent the vertices of an octahedron, a tetrahedron and another tetrahedron dual to the first tetrahedron. The last orbit $(011)_{D_3}$ represents the root system of the $D_3 \approx A_3$ Lie algebra and also constitutes the vertices of a cuboctahedron [Koca et al, 2007b]. The union of the orbits $(010)_{D_3}$ and $(001)_{D_3}$ represents the vertices of a cube.

Therefore the symmetry of the union of the orbits $(010)_{D_3} \cup (001)_{D_3}$ requires the Dynkin diagram symmetry $\gamma: \omega_1 \to \omega_1, \omega_2 \leftrightarrow \omega_3$, or equivalently, $\gamma: \alpha_1 \to \alpha_1, \alpha_2 \leftrightarrow \alpha_3$, which leads to the transformation on the imaginary quaternions $\gamma: e_1 \to e_1, e_2 \to e_2, e_3 \to -e_3$. In our formulation the Dynkin diagram symmetry operator reads $\gamma = [e_3, -e_3]^* = [e_3, e_3]$ which extends the Coxeter group $W(D_3)$ to the octahedral group $Aut(D_3) \approx W(D_3):C_2$, the automorphism group of the root system of $D_3$. The generators in (14) and $\gamma = [e_3, e_3]$ will lead to the quaternionic representation of the octahedral group of order 48

$$O_h \approx Aut(D_3) = \{[T, \pm\bar{T}] \cup [T', \pm\bar{T}']\}. \tag{19}$$

Note that from now on we are using the symbols of the sets representing their elements.
It is obvious that the maximal subgroups of the octahedral group $Aut(D_3) \approx O_h$ are the groups, each of which of order 24:

Chiral octahedral group : $O \approx \{[T, \bar{T}] \cup [T', \bar{T}']\}$,
Tetrahedral group $\quad : T_d \approx W(D_3) = \{[T, \bar{T}] \cup [T', -\bar{T}']\}$, $\quad$ (20)
Pyritohedral group $\quad : T_h = \{[T, \bar{T}] \cup [T, -\bar{T}]\}$.

Next, we would like to discuss the affine extension of the group $W(D_3)$.
Let us introduce the following definitions. For each root $\alpha$ and each integer $k$, we define an affine hyperplane (Humphreys, 1990)

$$H_{\alpha,k} := \{\lambda \in V \mid (\lambda, \alpha) = k\} \tag{21}$$

where $\lambda$ is an arbitrary vector in the Euclidean space $V$. Note that $H_{\alpha,k} = H_{-\alpha,-k}$ and the hyperplane $H_{\alpha,0} = H_\alpha$ coincides with the hyperplane passing through the origin orthogonal to the root $\alpha$. One can define the corresponding affine reflection as

$$r_{\alpha,k}(\lambda) := \lambda - \frac{2((\lambda,\alpha) - k)}{(\alpha,\alpha)}\alpha. \tag{22}$$

This reflection fixes $H_{\alpha,k}$ point wise and interchanges the vectors $0 \leftrightarrow \frac{2k\alpha}{(\alpha,\alpha)}$. Note also that the affine reflection can also be written as $t(\frac{2k\alpha}{(\alpha,\alpha)})r_\alpha$ where $t(\lambda)$ sends an arbitrary vector $\mu \to \mu + \lambda$.
The affine Coxeter group $W_a$ is generated by all affine reflections $r_{\alpha,k}$ where $\alpha$ is an arbitrary root and $k \in \mathbf{Z}$. It is possible to generate the group with a minimum number of generators. In addition to the generators $r_i$ $(i = 1, 2, ..., n)$, we introduce one more generator, usually denoted by $r_{\alpha_0} := r_0$, where $-\alpha_0 = \tilde{\alpha}$ is the highest root. The extended Coxeter-Dynkin diagram of $D_3$ is illustrated in Fig. 2. Here $\alpha_0$ represents the root of the extended Coxeter-Dynkin diagram. In the following we determine the affine group $W_a(D_3)$ and the associated lattices.

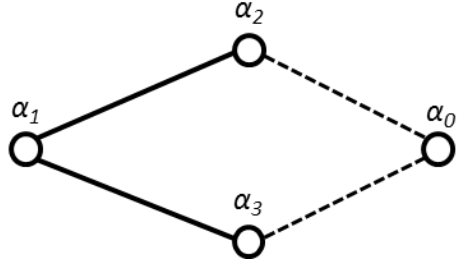

**Figure 2.** Extended Dynkin diagram of $D_3$.

The highest root of $D_3$ can be given in terms of quaternions as $\tilde{\alpha} = (\alpha_1 + \alpha_2 + \alpha_3) = e_1 + e_2$. We define the affine plane $H_{\tilde{\alpha},1} := \{\lambda \in V \mid (\lambda, \tilde{\alpha}) = 1\}$ and the reflection generator $r_0 = r_{\tilde{\alpha},1}(\lambda) := \lambda - \dfrac{2((\lambda, \tilde{\alpha}) - 1)}{(\tilde{\alpha}, \tilde{\alpha})} \tilde{\alpha}$. This generator represents the reflection with respect to the plane bisecting the highest root $\tilde{\alpha} = e_1 + e_2$.

Since $r_0$ involves a translation the affine Coxeter group $W_a(D_3)$ is generated by the four generators $\langle r_0, r_1, r_2, r_3 \rangle$. With the inclusion of the Dynkin diagram symmetry, the affine group will define a lattice with the octahedral point symmetry. If we take the generating vectors as the roots $\alpha_1, \alpha_2$ and $\alpha_3$ the affine group generate the fcc lattice (the root lattice $D_3$) where a general lattice vector is given by $p = b_1 \alpha_1 + b_2 \alpha_2 + b_3 \alpha_3$, $b_i \in \mathbf{Z}$. A vector in the root lattice can also be written in terms of the quaternionic imaginary units as $p = m_1 e_1 + m_2 e_2 + m_3 e_3$ where $\sum_{i=1}^{3} m_i = 2b_3 =$ even integer. This is the standard definition of the fcc lattice, namely, the sum of the integers $(m_1, m_2, m_3)$ in the orthonormal bases is equal to an even number. The orbit $(011)_{D_3}$, the set of root system of $D_3$, representing the centers of the edges of a cube represent the nonconventional unit cell of the fcc lattice (Ashcroft & Mermin, 1976). Moreover its dual polyhedron is the union of the orbits $(100)_{D_3} \cup \{(010)_{D_3} \cup (001)_{D_3}\}$ (Koca et al, 2010) which represents a rhombic dodecahedron as shown in Fig. 3. So the dual polyhedron of the non conventional cell of the fcc lattice is the Wigner-Seitz cell.

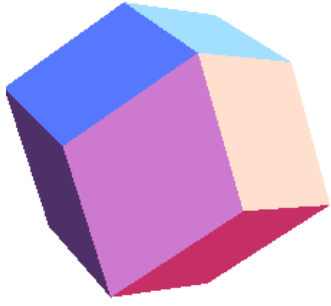

**Figure 3.** Rhombic dodecahedron, the Wigner–Seitz cell of the fcc lattice.

With the choice of the weight vectors in (17) the affine octahedral group generates the weight lattice $D_3^*$ which is the reciprocal lattice of the root lattice, that is, the bcc lattice. A general vector of the bcc lattice is given by $q = a_1\omega_1 + a_2\omega_2 + a_3\omega_3$, $a_i \in \mathbf{Z}$. In terms of quaternions it can be written as $q = n_1 e_1 + n_2 e_2 + n_3 e_3$, with either all $n_i$ integers or all half odd integers. Choosing the generating vector twice the weight vectors then an arbitrary bcc vector has the same form as above with $n_i$ are either all even integers or all odd integers. This is the usual definition of the bcc lattice in the literature. The unit cell, a cube including its center, can be represented by the union of the orbits $(010)_{D_3} \cup (001)_{D_3} \cup (000)_{D_3}$. The last orbit is the origin of the coordinate system representing the center of the cube of unit length. The Wigner-Seitz cell of the bcc lattice is the truncated octahedron which is represented by the orbit $\frac{1}{4}(111)_{D_3}$ consisting of 24 vertices given by

$$\frac{1}{4}(111)_{D_3} = \frac{1}{4}\{\pm 2e_1 \pm e_2, \pm 2e_2 \pm e_3, \pm 2e_3 \pm e_1, \pm e_1 \pm 2e_2, \pm e_2 \pm 2e_3, \pm e_3 \pm 2e_1\}. \qquad (23)$$

The factor 4 in the denominator is the Coxeter number of the group $W(D_3)$. The truncated octahedron constructed from the vertices (23) is depicted in Fig. 4.

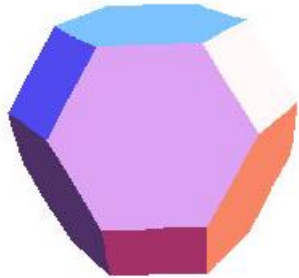

**Figure 4.** Truncated octahedron, the Wigner-Seitz cell of the bcc lattice.

## 3.2. Construction of the sc lattice with its affine Coxeter-Weyl group $W_a(B_3)$

The Coxeter-Dynkin diagram $B_3$ is illustrated in Fig. 5. The simple roots can be taken as $\alpha_1 = e_1 - e_2$, $\alpha_2 = e_2 - e_3$, $\alpha_3 = e_3$. Note that the last root is a short root and the angle between $\alpha_3$ and $\alpha_2$ is $135^0$.

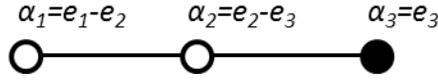

**Figure 5.** The Coxeter-Dynkin diagram $B_3$.

The generators of the Coxeter-Weyl group $W(B_3)$ are given by

$$r_1 = [\frac{1}{\sqrt{2}}(e_1 - e_2), \frac{1}{\sqrt{2}}(e_1 - e_2)],\ r_2 = [\frac{1}{\sqrt{2}}(e_2 - e_3), \frac{1}{\sqrt{2}}(e_2 - e_3)],\ r_3 = [e_3, e_3)]. \qquad (24)$$

They generate the octahedral group $W(B_3) \approx Aut(D_3) = \{[T, \pm\bar{T}] \cup [T', \pm\bar{T}']\}$. It is interesting to note that the first two generators in (24) are identical to the first two generators of the group $W(D_3)$ in (14) and the third generator is identical to the Dynkin diagram symmetry generator $\gamma$. This indicates that the third generator of the group $W(D_3)$ is redundant in the construction of $Aut(D_3)$. One can easily check that the third generator of the tetrahedral group $W(D_3)$ can be written as $\gamma r_2 \gamma$. The weight vectors of $B_3$ are given by

$$\omega_1 = e_1,\ \omega_2 = e_1 + e_2,\ \omega_3 = \frac{1}{2}(e_1 + e_2 + e_3). \qquad (25)$$

The orbits $(100)_{B_3}, (010)_{B_3}$ and $(001)_{B_3}$ respectively represent an octahedron, a cuboctahedron and a cube respectively. The truncated octahedron is represented by the orbit $(110)_{B_3}$. Note that the diagram $B_3$ consists of long roots and short roots of norms $\sqrt{2}$ and 1 respectively. The generators of the octahedral group $W(B_3) \approx O_h$ in (24) generate the root system consisting of the roots

$$(100)_{B_3} \cup (010)_{B_3} = \{\pm e_1, \pm e_2, \pm e_3\} \cup \{\pm e_1 \pm e_2, \pm e_2 \pm e_3, \pm e_3 \pm e_1\}. \qquad (26)$$

The short roots represent the centers of the faces (vertices of an octahedron) and the centers of the edges of a cube of 2 unit length. Before we proceed further we would like to discuss the structure of the octahedral group $W(B_3) \approx O_h$ in a little more detail. The proper subgroup (the chiral octahedral group) of the group $W(B_3) \approx O_h$ is generated by the rotation generators $a = r_1 r_2$, $b = r_1 r_3$ satisfying the generation relation $a^3 = b^4 = (ab)^2 = 1$. This relation is sufficient to prove that the chiral octahedral group can be generated by the generators $a$ and $b$.

A general vector of the root lattice then will be given by $p = b_1 \alpha_1 + b_2 \alpha_2 + b_3 \alpha_3, b_i \in \mathbf{Z}$. This indicates that a general vector of the lattice is the linear combinations of the imaginary quaternions $e_i$ with integer coefficients. It can also be represented as linear combinations of weight vectors $q = a_1 \omega_1 + a_2 \omega_2 + a_3 (2\omega_3)$, $a_i \in \mathbf{Z}$. The primitive cell of the lattice can be chosen as the cube with the vertices

$$0, e_i, e_i + e_j \, (i < j), e_i + e_j + e_k \, (i < j < k). \tag{27}$$

There are $2^3$ such cubes sharing the origin as a vertex. Denote by $V(0)$ the Voronoi cell around the origin. The vertices of the Voronoi polyhedron $V(0)$ can be determined as the intersection of the planes surrounding the origin. These planes are determined as the orbits of the fundamental weights

$$\frac{\omega_1}{2} = \frac{e_1}{2}, \frac{\omega_2}{2} = \frac{e_1 + e_2}{2}, \omega_3 = \frac{1}{2}(e_1 + e_2 + e_3). \tag{28}$$

The Voronoi polyhedron $V(0)$ (the Wigner-Seitz cell) is then the cube around the origin with the vertices

$$(001)_{B_3} = \frac{1}{2}(\pm e_1 \pm e_2 \pm e_3). \tag{29}$$

### 3.3. Construction of the pseudoicosahedron and its dual polyhedron from $D_3$ with the pyritohedral symmetry

Two dual platonic solids, the icosahedron and the dodecahedron possess the 5-fold symmetry in addition to 3-fold and 2-fold symmetries. The full symmetry of these polyhedra is the icosahedral symmetry of order 120 which can be derived from the noncrystallographic Coxeter diagram $H_3$. Its quaternionic description follows the same procedure studied in the previous sections.

Here one introduces the binary icosahedral subgroup of quaternions by $I = T \cup S$ with

$I = \langle p, q \rangle, \, p = \frac{1}{2}(e_1 + \tau e_2 + \sigma e_3), \, q = \frac{1}{2}(1 + e_1 + e_2 + e_3)$ and $\tau = \frac{1 + \sqrt{5}}{2}$ and $\sigma = \frac{1 - \sqrt{5}}{2}$

(Koca et al, 2007b) where $T$ is given in (5) and the set $S$ represents 96 remaining quaternions of the set I. The Coxeter group can be expressed as $W(H_3) = [I, \bar{I}] \cup [I, -\bar{I}]$. Since the icosahedral symmetry involves 5-fold symmetry it is not compatible with the crystallography which involves 2, 3, 4 and 6-fold symmetries only. However, there have been many discoveries in the quasicrystallography displaying the icosahedral symmetry (Shechtman et al., 1984). Here we will not pursue a discussion on the quasicrystal structures but rather on the 3D crystals with the pyritohedral symmetry which can be derived from the Coxeter-Dynkin diagram $D_3$. Here the pseudoicosahedron plays an essential role in understanding the crystallographic structures with the pyritohedral symmetry. We first note the fact that the pyritohedral symmetry $T_h = \{[T,\bar{T}] \cup [T,-\bar{T}]\}$ consists of only 3-fold and 2-fold symmetries and it is a maximal subgroup of both the octahedral group $W(B_3)$ and the icosahedral group $W(H_3)$.

### 3.4. Derivation of the vertices of the pseudoicosahedron from $D_3$ diagram

The rotation generators $r_1 r_2$ and $r_1 r_3$ of $D_3$ generate the subgroup $[T,\bar{T}] \approx A_4$ of order 12 representing the even permutations of the vertices of a tetrahedron. This is the rotational symmetry of a cube with stripes on its faces as shown in Fig. 6. The chiral tetrahedral group $[T,\bar{T}] \approx A_4$ involves 8 rotations by $120^0$ around the 4 diagonals of the cube, 3 rotations by $180^0$ around the $x, y$ and $z$ axes and the unit element. Pyrite crystals often occur in the shape of cubes with striated faces, octahedra and pyritohedra (a solid similar to dodecahedron but with non regular pentagonal faces), the shape of which will be discussed in this section.

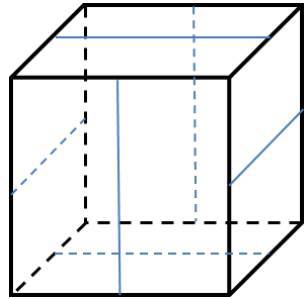

**Figure 6**. The cube with stripes on its faces.

Imposing the Dynkin diagram symmetry $\gamma$ the group is extended to the pyritohedral symmetry $T_h = A_4 \times C_2 = \langle r_1 r_2, r_1 r_3, \gamma \rangle$. Denote by a general vector of $D_3$ by $\lambda = a_1 \omega_1 + a_2 \omega_2 + a_3 \omega_3$ where $\omega_1, \omega_2$ and $\omega_3$ represent the weight vectors of (17). We note that the orbit of the pyritohedral group generated by the vector $\lambda$ with $a_1 = 1$, $a_2 = a_3 = 0$ is an octahedron. The orbit generated from $\lambda = \omega_2$ or $\lambda = \omega_3$ corresponds to a cube. Similarly the vector $\lambda = \omega_2 + \omega_3$ leads to the cuboctahedron under the pyritohedral symmetry.

From a general vector we can generate two equilateral triangles sharing the vertex $\lambda$. The vertices $\lambda$, $r_1 r_2 \lambda$ and $(r_1 r_2)^2 \lambda$ form an equilateral triangle with an edge length squared $a_1^2 + a_1 a_2 + a_2^2$. Similarly, the vertices $\lambda$, $r_1 r_3 \lambda$ and $(r_1 r_3)^2 \lambda$ form the second triangle with the edge length squared is $a_1^2 + a_1a_3 + a_3^2$. If one draws a line between the vertices $\lambda$ and $r_2r_3\lambda$ its length squared will be $a_2^2 + a_3^2$. These vectors constitute five vertices $r_1r_2\lambda$, $(r_1r_2)^2\lambda$, $r_1r_3\lambda$, $(r_1r_3)^2\lambda$ and $r_2r_3\lambda$ surrounding the vertex $\lambda$ as shown in Fig. 7. If we impose all the edge lengths take the same value then one obtains five equilateral triangles around one vertex. This would lead to the set of equations

$$a_1^2 + a_1a_2 + a_2^2 = a_1^2 + a_1a_3 + a_3^2 = a_2^2 + a_3^2. \tag{30}$$

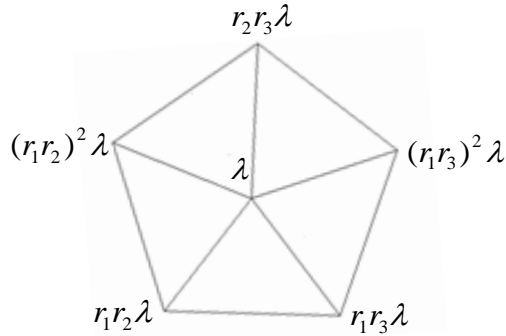

**Figure 7.** Five triangles meeting at one vertex.

Let us assume that the general vector $\lambda$ is invariant under the Dynkin diagram symmetry, that is, $\gamma\lambda = \lambda$. This would imply that $a_2 = a_3$ and a general vector can be written as

$$\lambda = a_1(\omega_1 + x(\omega_2 + \omega_3)) = a_1((1+x)e_1 + xe_2) \tag{31}$$

where $x = \dfrac{a_2}{a_1} = \dfrac{a_3}{a_1}$ is the parameter which could be computed from (30). Factoring (30) by $a_1$ one obtains the equation $x^2 - x - 1 = 0$. The solutions of this equation are $\tau = \dfrac{1+\sqrt{5}}{2}$ and $\sigma = \dfrac{1-\sqrt{5}}{2}$. The action of the pyritohedral group on the vector given in (31) apart from the scale factor $a_1$ will generate the set of 12 vectors:

$$\pm(1+x)e_1 \pm xe_2, \pm(1+x)e_2 \pm xe_3, \pm(1+x)e_3 \pm xe_1. \tag{32}$$

Substituting $\tau$ and $\sigma$ respectively for $x$ we obtain two sets of 12 vertices as:

$$\tau\{\pm\tau e_1 \pm e_2, \pm\tau e_2 \pm e_3, \pm\tau e_3 \pm e_1\}, \tag{33a}$$

$$\sigma\{\pm\sigma e_1 \pm e_2, \pm\sigma e_2 \pm e_3, \pm\sigma e_3 \pm e_1\}. \tag{33b}$$

These sets of vertices represent two mirror images of an icosahedron albeit a scale factor difference. Multiplying the vertices in (33b) by $\tau^3$ one obtains the following set of quaternions

$$\tau\{\pm e_1 \pm \tau e_2, \pm e_2 \pm \tau e_3, \pm e_3 \pm \tau e_1\}. \tag{34}$$

The icosahedron represented by (33a) is depicted in Fig. 8.

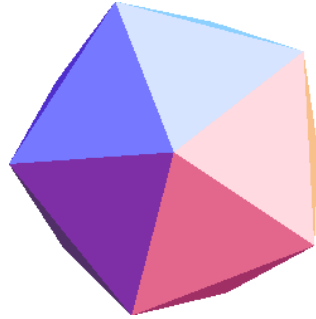

**Figure 8.** The icosahedron.

The vertices of the dual of the icosahedron, say, the set of vectors of (33a) can be determined as (Koca et al, 2011)

$$\frac{1}{2}\{\pm\sigma e_1 \pm \tau e_2, \pm\sigma e_2 \pm \tau e_3, \pm\sigma e_3 \pm \tau e_1\}, \tag{35a}$$

$$\frac{1}{2}(\pm e_1 \pm e_2 \pm e_3). \tag{35b}$$

The 20 vertices in (35a-b) represent a dodecahedron as shown in Fig. 9. Its mirror image can be obtained by replacing $\sigma \leftrightarrow \tau$.

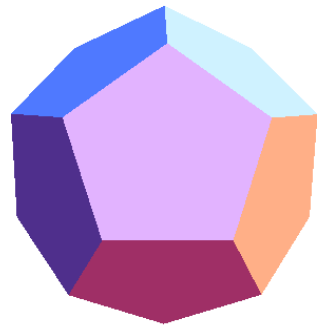

**Figure 9.** The dodecahedron, dual of the icosahedron.

The vectors in (35b) represent the vertices of a cube which is invariant under the pyritohedral symmetry. Similarly the other 12 vertices of (35a) form another orbit under the pyritohedral symmetry. One may now wonder what solid do they represent? Actually they can be obtained from (32) by substituting $x = -\tau$. We note that when $x = -\tau$ the equation in (30) is violated and we obtain two different edge lengths. This follows from the relation $a_1^2 + a_1 a_2 + a_2^2 = a_2^2 + a_2 a_3 + a_3^2 \neq a_2^2 + a_3^2$ leading to two classes of triangles surrounding the vertex $\lambda$: two are equilateral triangles and the other three are isosceles triangles. For any general value of the parameter $x \neq 0$, $x \neq -\frac{1}{2}$ and $x \neq -1$ the set of vertices in (32) represents a pseudoicosahedron consisting of 8 equilateral triangles and 12 isosceles triangles.

### 3.5. Derivation of the vertices of the pyritohedron from the pseudoicosahedron

We can compute the dual of the pseudoicosahedron of (30) for an arbitrary value of $x$. (Since the dual of the icosahedron is the dodecahedron the dual of the pseudoicosahedron could be called pseudododecahedron instead of pyritohedron!). This can be achieved by determining the vectors normal to the faces of the pseudoicosahedron of (30). From Fig. 8 the center of the triangle with the vertices $\lambda$, $r_1 r_2 \lambda$, $(r_1 r_2)^2 \lambda$ can be taken as $\omega_3$ because $r_1 r_2$ is a rotation around $\omega_3$ for $r_1 r_2 \omega_3 = \omega_3$. Similarly the center of the other equilateral triangle can be taken as $\omega_2$. As we have pointed out before, the vertices generated by the pyritohedral group from either $\omega_2$ or $\omega_3$ would lead to the vertices of a cube given in (35b). The vectors normal to the isosceles triangles can be computed as:

$$\begin{aligned} b_1 &= e_1 + (1+x)e_2, \\ b_4 &= (1+x)e_1 + e_3, \\ b_5 &= (1+x)e_1 - e_3. \end{aligned} \qquad (36)$$

Note that the vectors $b_1, b_4$ and $b_5$ are in the same orbit under the pyritohedral group. They form a plane orthogonal to the vector $d = (1+x)e_1 + xe_2$ to which $\omega_3 - \omega_2 = e_3$ is also orthogonal. Were these five vertices to determine the same plane then one can show that $((\rho b_1 - \omega_2), d) = 0$ is necessary and sufficient. From this relation one obtains $\rho = \frac{1+2x}{2(1+x)^2}$. Then the vectors $\rho b_1, \rho b_4, \rho b_5, \omega_2$ and $\omega_3$ determine a pentagon, non-regular in general. Applying the pyritohedral group on these vertices one obtains the set of vectors in two orbits, one set with 12 vectors and the other with 8 vectors:

$$\frac{(1+2x)}{2(1+x)^2}\{\pm e_1 \pm (1+x)e_2, \pm e_2 \pm (1+x)e_3, \pm e_3 \pm (1+x)e_1\},$$
$$\frac{1}{2}(\pm e_1 \pm e_2 \pm e_3). \tag{37}$$

These are the vertices of the dual solid (pseudo dodecahedron) of the pseudoicosahedron of (32). It is generally called pyritohedron which is made of 12 pentagonal faces. A simpler form of this can be obtained by defining the parameter $h = \dfrac{x}{x+1}$. Dropping an overall factor $1/2$ then (37) will read

$$\{\pm(1-h^2)e_1 \pm (1+h)e_2, \pm(1-h^2)e_2 \pm (1+h)e_3, \pm(1-h^2)e_3 \pm (1+h)e_1\},$$
$$(\pm e_1 \pm e_2 \pm e_3). \tag{38}$$

For a general $x$ the union of two pseudoicosahedra

$$\pm(1+x)e_1 \pm xe_2, \pm(1+x)e_2 \pm xe_3, \pm(1+x)e_3 \pm xe_1,$$
$$\pm xe_1 \pm (1+x)e_2, \pm xe_2 \pm (1+x)e_3, \pm xe_3 \pm (1+x)e_1 \tag{39}$$

represents the vertices of a non-regular truncated octahedron which can be derived as the orbit of the Coxeter group $W(B_3)$ denoted by $a_1(1,x,0)_{B_3}$. This belongs to the simple cubic lattice if $a_1$ and $a_1 x = a_2$ are integers which implies that $x$ should be a rational number. In other words as long as $a_1$ and $a_2$ are integers the pseudoicosahedron with the vertices

$$\pm(a_1+a_2)e_1 \pm a_2 e_2, \pm(a_1+a_2)e_2 \pm a_2 e_3, \pm(a_1+a_2)e_3 \pm a_2 e_1 \tag{40}$$

can be embedded in a simple cubic lattice. Of course its mirror image obtained by the interchange $e_1 \leftrightarrow e_2$ in (40) also represents a pseudoicosahedron embedded in the simple cubic lattice.

### 3.6. Fibonacci sequence of pseudoicosahedra

Fibonacci sequence consists of the numbers in the following integer sequence: 1,1,2,3,5,8,13,21,…which can be put in a recurrence relation $F_n = F_{n-1} + F_{n-2}$ with values $F_1 = F_2 = 1$. The general term of the Fibonacci sequence can be written as

$$F_n = \frac{\tau^n - \sigma^n}{\tau - \sigma} . \tag{41}$$

The ratio of Fibonacci sequence numbers converges to the golden ratio $\tau$:

$$\lim_{n \to \infty} \frac{F_{n+1}}{F_n} = \tau \qquad (42)$$

Let us consider the interval $1 \leq x \leq 2$ and define the sequence that the parameter $x$ takes the rational numbers $x_n = \frac{F_{n+1}}{F_n}$ such as

$$1, \frac{2}{1}, \frac{3}{2}, \frac{5}{3}, \frac{8}{5}, \frac{13}{8}, \frac{21}{13}, \frac{34}{21}, \frac{55}{34}, \frac{89}{55}, \ldots \qquad (43)$$

The sequence of (43) takes values in the interval $1 \leq x \leq 2$ and approaches to the golden ratio $\tau$ in the limit. This leads to an infinite sequence of pseudoicosahedra in the simple cubic lattice with the appropriate choice of the integer parameters $a_1$. For example, choosing $a_1 = 3$ for $x_4 = \frac{5}{3}$ then the vertices of the pseudoicosahedron would read: $\pm 8e_1 \pm 5e_2, \pm 8e_2 \pm 5e_3, \pm 8e_3 \pm 5e_1$. For the mirror pseudoicosahedron just change $e_1 \leftrightarrow e_2$. Similar consideration can be made for the pyritohedron.

It is also interesting to note that the centers of the edges of the pseudoicosahedron of (32) represent 30 vertices of a pseudo icosidodecahedron given in the form of two orbits under the pytitohedral group by:

$$a_1 \{\pm xe_1 \pm (1+x)e_2 \pm (1+2x)e_3, \pm (1+x)e_1 \pm (1+2x)e_2 \pm xe_3, \pm (1+2x)e_1 \pm xe_2 \pm (1+x)e_3\},$$

$$a_1(1+x)\{\pm e_1, \pm e_2, \pm e_3\}. \qquad (44)$$

In the limit $x \to \tau$ the vertices in (44) represent a regular icosidodecahedron with two regular pentagons and two equilateral triangles meeting at every vertex as shown in Fig. 10(a). Otherwise the regular pentagons are replaced by irregular pentagons (Fig. 10(b)).

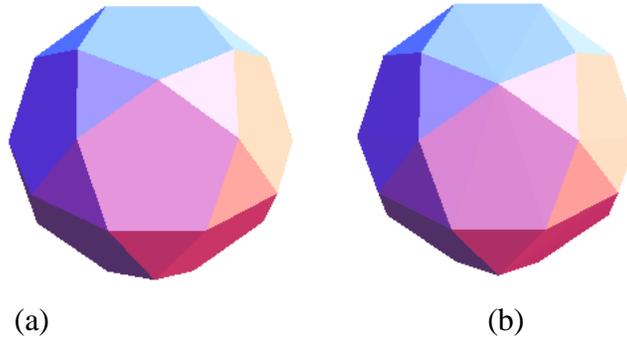

(a)          (b)

**Figure 10.**(a) An icosidodecahedron with regular pentagons and equilateral triangular faces, (b) A pseudo icosidodecahedron obtained from (44) for $x = \frac{3}{2}$.

## 4. Pseudo snub 24-cell derived from $D_4$ diagram

Let us recall that the pyritohedral group $[T,\bar{T}]\cup[T,\bar{T}]^*$ is derived from the Coxeter-Dynkin diagram $D_3$ by the rotation generators and the Dynkin diagram symmetry. The straightforward generalization of this group to 4D is to start with the rotation generators of $D_4$ and impose the Dynkin diagram symmetry. We will see that the group generated from the diagram $D_4$ is nothing other than the group $[T,T]\cup[T,T]^*$ of order 576. It represents the symmetry of the snub 24-cell (Coxeter,1973; Koca et al. 2007a). Let us define the subsets of the quaternions $T=V_0\cup V_+\cup V_-$ and $T'=V_1\cup V_2\cup V_3$ as follows:

$$V_0=\{\pm 1,\pm e_1,\pm e_2,\pm e_3\},$$
$$V_+=\frac{1}{2}(\pm 1\pm e_1\pm e_2\pm e_3),\text{ even number of (-) sign,} \quad (45)$$
$$V_-=\frac{1}{2}(\pm 1\pm e_1\pm e_2\pm e_3),\text{ odd number of (-) sign,}$$

$$V_1=\{\frac{1}{\sqrt{2}}(\pm 1\pm e_1),\frac{1}{\sqrt{2}}(\pm e_2\pm e_3)\},$$
$$V_2=\{\frac{1}{\sqrt{2}}(\pm 1\pm e_2),\frac{1}{\sqrt{2}}(\pm e_3\pm e_1)\}, \quad (46)$$
$$V_3=\{\frac{1}{\sqrt{2}}(\pm 1\pm e_3),\frac{1}{\sqrt{2}}(\pm e_1\pm e_2)\}.$$

They satisfy the multiplication table shown in Table 1.

Table 1. The multiplication table of the sets of quaternions $V_0,V_+,V_-,V_1,V_2,V_3$.

|  | $V_0$ | $V_+$ | $V_-$ | $V_1$ | $V_2$ | $V_3$ |
|---|---|---|---|---|---|---|
| $V_0$ | $V_0$ | $V_+$ | $V_-$ | $V_1$ | $V_2$ | $V_3$ |
| $V_+$ | $V_+$ | $V_-$ | $V_0$ | $V_3$ | $V_1$ | $V_2$ |
| $V_-$ | $V_-$ | $V_0$ | $V_+$ | $V_2$ | $V_3$ | $V_1$ |
| $V_1$ | $V_1$ | $V_2$ | $V_3$ | $V_0$ | $V_+$ | $V_-$ |
| $V_2$ | $V_2$ | $V_3$ | $V_1$ | $V_-$ | $V_0$ | $V_+$ |
| $V_3$ | $V_3$ | $V_1$ | $V_2$ | $V_+$ | $V_-$ | $V_0$ |

These subsets are useful to denote the Coxeter Weyl group $W(D_4)$ in a compact form. The Coxeter-Dynkin diagram $D_4$ is depicted in Fig. 11 with the simple roots,

$$\alpha_1 = 1-e_1,\ \alpha_2 = e_1-e_2,\ \alpha_3 = e_2-e_3,\ \alpha_4 = e_2+e_3. \tag{47}$$

The corresponding weights are determined as

$$\omega_1 = 1,\ \omega_2 = 1+e_1,\ \omega_3 = \frac{1}{2}(1+e_1+e_2-e_3),\ \omega_3 = \frac{1}{2}(1+e_1+e_2+e_3). \tag{48}$$

Note that $\frac{\alpha_i}{\sqrt{2}} \in T'$, $i = 1,2,3,4$; $\frac{\omega_2}{\sqrt{2}} \in T'$; $\omega_j \in T$, j=1,3,4.

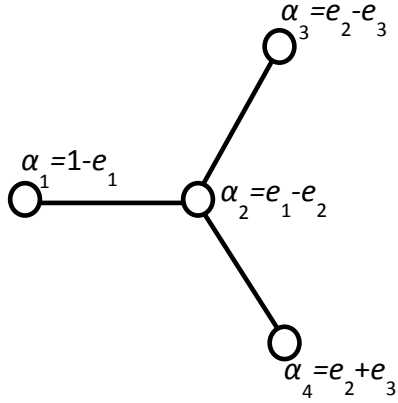

**Figure 11.** The Coxeter-Dynkin diagram $D_4$.

In terms of quaternionic simple roots the group generators of $W(D_4)$ can be written as

$$\begin{aligned}
r_1 &= [\frac{1}{\sqrt{2}}(1-e_1),\ -\frac{1}{\sqrt{2}}(1-e_1)]^*, \\
r_2 &= [\frac{1}{\sqrt{2}}(e_1-e_2),\ -\frac{1}{\sqrt{2}}(e_1-e_2)]^*, \\
r_3 &= [\frac{1}{\sqrt{2}}(e_2-e_3),\ -\frac{1}{\sqrt{2}}(e_2-e_3)]^*, \\
r_4 &= [\frac{1}{\sqrt{2}}(e_2+e_3),\ -\frac{1}{\sqrt{2}}(e_2+e_3)]^*.
\end{aligned} \tag{49}$$

They generate the Coxeter-Weyl group $W(D_4)$ of order 192 which can be represented by the set [Koca et al., 2001]

$$W(D_4) = \{[V_0,V_0] \cup [V_+,V_-] \cup [V_-,V_+]\} + \{[V_1,V_1]^* \cup [V_2,V_2]^* \cup [V_3,V_3]^*\}. \tag{50}$$

Note that the subset of the Coxeter-Weyl group

$$W(D_4)/C_2 = \{[V_0, V_0] \cup [V_+, V_-] \cup [V_-, V_+]\} \qquad (51)$$

represents the proper subgroup and can be directly generated by the rotation generators $r_2 r_1$, $r_2 r_3$, $r_2 r_4$. Let us impose the Dynkin diagram symmetry which is the permutation group $S_3$ of the simple roots $\alpha_1$, $\alpha_3$ and $\alpha_4$. The permutation group is of order 6 and can be generated, for example, by two generators $[p, q] \in [V_+, V_+]$ with $[p, q]^3 = [1, 1]$ and $[e_3, -e_3]^* \in [V_0, V_0]^*$ where $p = \frac{1}{2}(1 + e_1 - e_2 - e_3)$ and $q = \frac{1}{2}(1 - e_1 + e_2 - e_3)$. The group $W(D_4)/C_2$ is invariant under conjugation of the group $S_3$. We first note that the extension of the group of (51) by the cyclic group of order 3 generated by the generator $[p, q]$ is a group of order 288 which can be denoted by

$$\left(W(D_4)/C_2\right) : C_3 = [T, T]. \qquad (52)$$

The extension by the full permutation group $S_3$ [ see Armstrong, 1988 for a discussion of the permutation groups] is given as the semi-direct product of two groups as:

$$\left(W(D_4)/C_2\right) : S_3 = \{[T, T] \cup [T, T]^*\} . \qquad (53)$$

As we will see in the next section this is the symmetry group of the snub 24-cell as well as that of any pseudo snub 24-cell.

### 4.1 Construction of the vertices of the pseudo snub-24 cell

The affine extension of the Coxeter group $W(D_4)$ can be made by introducing the affine generator $r_0 \Lambda = \Lambda - \frac{2[(\Lambda, \tilde{\alpha}) - 1]}{(\tilde{\alpha}, \tilde{\alpha})} \tilde{\alpha}$ where $\tilde{\alpha} = 1 + e_1$ is the highest root. See, for instance, [Slansky, 1981] for the definition of the highest root. Then the affine Coxeter group can be generated by five generators including $r_0$, $W_a(D_4) = \langle r_0, r_1, r_2, r_3, r_4 \rangle$ where $r_0$ represents the reflection with respect to the hyperplane bisecting the line from the origin to the highest root $\tilde{\alpha} = 1 + e_1$.

Applying the group $W_a(D_4)$ on any simple root one can generate the root lattice where an arbitrary lattice vector can be written as

$$\Lambda = b_1 \alpha_1 + b_2 \alpha_2 + b_3 \alpha_3 + b_4 \alpha_4, \ b_i \in \mathbf{Z}, \ i = 1, 2, 3, 4. \qquad (54)$$

It can be expressed in terms of quaternionic units by

$$\Lambda = m_0 + m_1 e_1 + m_2 e_2 + m_3 e_3, \ m_i \in \mathbf{Z}, \ i = 0,1,2,3 \tag{55}$$

where $\sum_{i=0}^{3} m_i = 2b_4 =$ even integer. This shows that the vector $\Lambda$ determines the root lattice $D_4$. In this section we are interested in the derivation of the vertices of the polytope ( pseudo snub 24-cell) in terms of the root lattice vectors of $D_4$. The $D_4$ is a self-dual lattice [Conway & Sloane, 1988] so that we can also express the dual lattice vectors in terms of the weight vectors as

$$\Lambda = a_1 \omega_1 + a_2 \omega_2 + a_3 \omega_3 + a_4 \omega_4, \ a_i \in \mathbf{Z}, \ i = 1,2,3,4. \tag{56}$$

Let us follow the argument that we used for the Dynkin diagram symmetry for $D_3$; we impose the Dynkin diagram symmetry $S_3$ on the vector $\Lambda$, that is, $S_3 \Lambda = \Lambda$. The group $S_3$ permutes the weight vectors $\omega_1, \omega_3$ and $\omega_4$ but leaves $\omega_2$ invariant. Similarly, the generators $r_1, r_3$ and $r_4$ are permuted and the generator $r_2$ is left invariant under the group conjugation by the group $S_3$. This implies that the vector $\Lambda$ takes the form

$$\Lambda = a_2 \omega_2 + a_1 (\omega_1 + \omega_3 + \omega_4), \ a_i \in \mathbf{Z}, \ i = 1,2. \tag{57}$$

If we factorize by $a_2$ and define the rational number $x = \dfrac{a_1}{a_2}$ the vector $\Lambda$ reads in terms of quaternions

$$\Lambda = a_2 [(1+2x) + (1+x)e_1 + xe_2]. \tag{58}$$

Note that the sum of the coefficients of the quaternionic units is an even integer as we mentioned earlier. For values of $x = -1, -\dfrac{1}{2}, 0$ the vector $\Lambda$ belongs to the set of quaternions $\Lambda \in T'$ which is known to be the polytope 24-cell. For these particular values of $x$ the group $\left(W(D_4)/C_2\right) : S_3 = \{[T,T] \cup [T,T]^*\}$ generates 24 vertices of $T'$. The polytope 24-cell has 24 vertices, 96 edges, 96 triangular faces and 24 octahedral cells. 6 octahedra meet at one vertex. It is self dual in the sense that the dual of $T'$ is the set of quaternions $T$. The polytope 24-cell constitutes the unit cell of the $D_4$ lattice. Its Voronoi cell is also the 24-cell. If $T$ represents the unit cell of the root lattice then the set $\dfrac{T'}{\sqrt{2}}$ represents its Voronoi cell. Applying the group elements represented by (53) on the vector in (57) we obtain 96 vertices as the orbit of the group $\left(W(D_4)/C_2\right) : S_3$ as expected for $\dfrac{576}{6} = 96$. We list 96 vertices dropping the overall factor $a_2$ as follows:

$$S(x) =$$
$$\{\pm(1+2x) \pm (1+x)e_1 \pm xe_2, \pm x \pm (1+2x)e_1 \pm (1+x)e_2, \pm(1+x) \pm xe_1 \pm (1+2x)e_2,$$
$$\pm(1+2x) \pm (1+x)e_2 \pm xe_3, \pm x \pm (1+2x)e_2 \pm (1+x)e_3, \pm(1+x) \pm xe_2 \pm (1+2x)e_3,$$
$$\pm(1+2x) \pm xe_1 \pm (1+x)e_3, \pm x \pm (1+x)e_1 \pm (1+2x)e_3, \pm(1+x) \pm (1+2x)e_1 \pm xe_3,$$
$$\pm(1+2x)e_1 \pm xe_2 \pm (1+x)e_3, \pm xe_1 \pm (1+x)e_2 \pm (1+2x)e_3, \pm(1+x)e_1 \pm (1+2x)e_2 \pm xe_3\}.$$
(59)

The mirror image of the pseudo snub 24-cell in (59) can be obtained by applying any reflection generator of $D_4$. For example if one applies $r_2$ on $S(x)$ which interchanges $e_1 \leftrightarrow e_2$ and leaves the other quaternionic units intact then one obtains the mirror image of the pseudo snub 24-cell in (59). Both the pseudo snub 24-cell and its mirror image lie in the $D_4$ lattice. It is clear that for $x = -1, -\frac{1}{2}, 0$ the vertices of (59) reduces to the set $T'$. Excluding these values of $x$ the set of vertices represent a pseudo snub 24-cell. Only in the limit $x \to \frac{1 \pm \sqrt{5}}{2}$ the vertices in (59) represent the snub 24 cell [Koca et al., 2012]. Since in this case $x$ is not a rational number the vertices do not belong to the lattice $D_4$. We shall prove in Section 5 that the snub 24-cell can be obtained as the limit of the Fibonacci sequence of the pseudo snub 24-cells.

It is well known that every vertex of the snub 24-cell is surrounded by three icosahedra and five tetrahedra. In what follows we shall prove that the facets of the pseudo snub 24-cell consist of pseudo icosahedra, tetrahedra and triangular pyramids. In the pseudo snub 24-cell three pseudoicosahedra, one tetrahedron and four triangular pyramids meet at the same vertex. Now we discuss the details of this structure. It is evident from the diagram $D_4$ that each of the following set of rotation generators $(r_1r_2, r_2r_3), (r_3r_2, r_2r_4)$ and $(r_4r_2, r_2r_1)$ generate a proper subgroup of the tetrahedral group of order 12. Let us discuss how one of these groups act on the vertex $\Lambda$. Let us take the group $\langle r_1r_2, r_2r_3 \rangle = [T, \bar{\omega}_4 \bar{T} \omega_4]$ which implies that the group consist of 12 elements leaving the weight vector $\omega_4 \in T$ invariant. The group generators transform quaternionic units as follows:

$$r_1r_2 : 1 \to e_1 \to e_2 \to 1,\ e_3 \to e_3$$
$$r_2r_3 : 1 \to 1,\ e_1 \to e_2 \to e_3 \to e_1.$$
(60)

The 12 vertices generated by the group can be written as:

$$(1+2x)+(1+x)e_1+xe_2, x+(1+2x)e_1+(1+x)e_2,(1+x)+xe_1+(1+2x)e_2,$$
$$(1+2x)+(1+x)e_2+xe_3, x+(1+2x)e_2+(1+x)e_3,(1+x)+xe_2+(1+2x)e_3,$$
$$(1+2x)+xe_1+(1+x)e_3, x+(1+x)e_1+(1+2x)e_3,(1+x)+(1+2x)e_1+xe_3, \quad (61)$$
$$(1+2x)e_1+xe_2+(1+x)e_3, xe_1+(1+x)e_2+(1+2x)e_3,(1+x)e_1+(1+2x)e_2+xe_3.$$

Since the vector $\omega_4$ is left invariant by the generators in (60), the center of the pseudo icosahedron of (61) can be taken as $\omega_4$ up to a scale factor. One may check that the set in (61) is also left invariant under the group element $[\omega_4, \omega_4]^*$. Therefore the set (61) is invariant under the larger group $T_h \approx [T, \bar{\omega}_4 \bar{T} \omega_4] \cup [T, \omega_4 \bar{T} \omega_4]^*$ of order 24 isomorphic to the pyritohedral group. If we define a new set of unit quaternions $p_0 = \omega_4, p_1 = e_1 p_0, p_2 = e_2 p_0, p_3 = e_3 p_0$ and express the vectors in (61) in terms of the new set of quaternions, apart from an overall component proportional to $p_0$ the rest would read the same equation in (32) where $e_1, e_2$ and $e_3$ are replaced by $p_1, p_2$ and $p_3$ respectively. This proves that the set of vertices in (61) represent a pseudo icosahedron. When the set of generators $(r_3 r_2, r_2 r_4)$ and $(r_4 r_2, r_2 r_1)$ respectively applied to the vector $\Lambda$ one generates two more pseudo icosahedra with the centers represented by $\omega_1$ and $\omega_3$ respectively. The groups generating the vertices of the second and the third pseudo icosahedra can be written respectively as $[T, \bar{\omega}_1 \bar{T} \omega_1] \cup [T, \omega_1 \bar{T} \omega_1]^*$, $\omega_1 \in T$ and $[T, \bar{\omega}_3 \bar{T} \omega_3] \cup [T, \omega_3 \bar{T} \omega_3]^*$, $\omega_3 \in T$. These groups are isomorphic to the pyritohedral group in (20). When $\omega_1 = 1$ is substituted above we get exactly the same group elements in (20) when the basis vectors are chosen as $e_1, e_2$ and $e_3$ in 3D. Note that $[1, -1] \in [T, T]$ is also an element of the group $[T, T]$ which commutes with all elements and sends a quaternion to its negative $q \to -q$. By adjoining the generator $[1, -1]$ to the pyritohedral group one obtains a larger group $T_h \times C_2 \approx [T, \pm t\bar{T}t] \cup [T, \pm t\bar{T}t]^*$, $t \in T$ of order 48 leaving the vector $\pm t$ invariant which can be embedded in the group $\left( W(D_4)/C_2 \right) : S_3$ 12 different ways proving that the number of pseudo icosahedra generated by these groups is 24. The group $[T, T] \cup [T, T]^*$ also leaves the set of quaternions $T'$ invariant since $TT' \in T'$ and $T'T \in T'$. The maximal subgroup which leaves $\pm t' \in T'$ invariant can be written as $T_d \times C_2 \approx [T, \pm t'\bar{T}t'] \cup [T, \pm t'\bar{T}t']^*$, $t' \in T'$.

Now we continue to discuss the polyhedral facets having $\Lambda$ as a vertex. Let us consider the following five sets of rotational generators obtained from the generators of the Coxeter-Weyl group $W(D_4)$:

$$(r_1 r_3, r_3 r_4, r_4 r_1), \ (r_2 r_1, r_2 r_3, r_2 r_4), \ (r_1 r_2, r_1 r_3, r_1 r_4), \ (r_3 r_1, r_3 r_2, r_3 r_4), \ (r_4 r_1, r_4 r_2, r_4 r_3). \quad (62)$$

The first two sets of generators are invariant under the conjugation of the permutation group $S_3$ but the next three sets of generators are permuted among each other. Let us determine the vertices of the polyhedra under the action of five sets.

1. The set of vertices

$$\Lambda = (1+2x)+(1+x)e_1 + xe_2,$$
$$r_1 r_3 \Lambda = (1+x)+(1+2x)e_1 + xe_3,$$
$$r_3 r_4 \Lambda = (1+2x)+(1+x)e_1 - xe_2, \qquad (63)$$
$$r_4 r_1 \Lambda = (1+x)+(1+2x)e_1 - xe_3,$$

determines a tetrahedron of edge length $2x$ as shown in Fig. 12. Its center can be represented by $p(1) = \omega_2 = 1+e_1$ up to some scale factor. Since the Dynkin diagram symmetry $S_3$ also leaves $\omega_2 = 1+e_1$ invariant the group $C_2 \times C_2$ generated by the generators $(r_1 r_3, r_3 r_4, r_4 r_1)$ can be extended by the $S_3$ symmetry to a group $[T, \bar{\omega}_2 \bar{T} \omega_2] \cup [T, \omega_2 \bar{T} \omega_2]^*$ of order 24 isomorphic to the tetrahedral group in (20). Since $S_3 \Lambda = \Lambda$ the group $T_d \approx [T, \bar{\omega}_2 \bar{T} \omega_2] \cup [T, \omega_2 \bar{T} \omega_2]^*$, $\omega_2 \in T'$ leaves the vertices of tetrahedron in (63) invariant. Note that this is not pyritohedral symmetry. This tells us that the number of tetrahedra generated by the conjugate tetrahedral groups is also 24. Extension of the group by the generator $[1,-1]$ leads to the group $[T, \pm\bar{\omega}_2 \bar{T} \omega_2] \cup [T, \pm\omega_2 \bar{T} \omega_2]^*$, $\pm\omega_2 \in T'$ that is isomorphic to the octahedral group in (19).

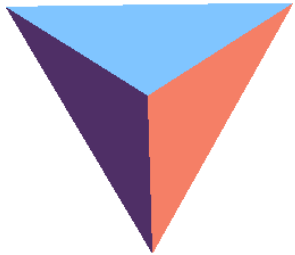

**Figure 12.** Tetrahedron with edges $2x$ ($x=1$).

2. The set of vertices

$$\Lambda = (1+2x)+(1+x)e_1 + xe_2,$$
$$r_2 r_1 \Lambda = 1+x+xe_1+(1+2x)e_2,$$
$$r_2 r_3 \Lambda = (1+2x)+(1+x)e_1 + xe_3, \qquad (64)$$
$$r_2 r_4 \Lambda = (1+2x)+(1+x)e_2 - xe_3,$$

represent a triangular pyramid with a base of equilateral triangle with sides $2x$ and the other edges of length $\sqrt{2(x^2+x+1)}$. The hyperplane determined by these four vertices in (64) is orthogonal to the vector $p(2)=(2+x)+e_1+(1+x)e_2$.

The group generated by these generators extended by the group $S_3$ is the full group of symmetry $\left(W(D_4)/C_2\right):S_3$.

3. The set of vertices
$$\begin{aligned}\Lambda &= (1+2x)+(1+x)e_1+xe_2,\\ r_1r_2\Lambda &= x+(1+2x)e_1+(1+x)e_2,\\ r_1r_3\Lambda &= (1+x)+(1+2x)e_1+xe_3,\\ r_1r_4\Lambda &= (1+x)+(1+2x)e_1-xe_3,\end{aligned} \quad (65)$$

defines another triangular pyramid with the same edge lengths as above. The vector orthogonal to the hyperplane determined by the vertices in (65) is $p(3)=(1+x)+(2+x)e_1+e_2$.

4. The set of vertices
$$\begin{aligned}\Lambda &= (1+2x)+(1+x)e_1+xe_2,\\ r_3r_1\Lambda &= (1+x)+(1+2x)e_1+xe_3,\\ r_3r_2\Lambda &= (1+2x)+xe_1+(1+x)e_3,\\ r_3r_4\Lambda &= (1+2x)+(1+x)e_1-xe_2,\end{aligned} \quad (66)$$

also determines a triangular pyramid as above. The vector which is orthogonal to the hyperplane determined by the vectors of (66) can be computed as $p(4)=(2+x)+(1+x)e_1+e_3$.

5. The set of vertices
$$\begin{aligned}\Lambda &= (1+2x)+(1+x)e_1+xe_2,\\ r_4r_1\Lambda &= (1+x)+(1+2x)e_1-xe_3,\\ r_4r_2\Lambda &= (1+2x)+xe_1-(1+x)e_3,\\ r_4r_3\Lambda &= (1+2x)+(1+x)e_1-xe_2,\end{aligned} \quad (67)$$

defines another triangular pyramid with the same edge lengths as above. The vector $p(5)=(2+x)+(1+x)e_1-e_3$ is orthogonal to the hyperplane of the vectors in (67).

Since the groups generated by the generators in (64-67) is the full symmetry group, all vectors $p(i)$, $i = 2, 3, 4, 5$ lie in the same orbit of the group $\left(W(D_4)/C_2\right):S_3$.

Below we will list these five vectors representing the centers of the above polyhedra expressed in terms of the weight vectors $\omega_i, i = 1, 2, 3, 4$:

$$\begin{aligned}
p(1) &= \omega_2, \\
p(2) &= (1+x)(\omega_1 + \omega_3 + \omega_4) - x\omega_2, \\
p(3) &= -\omega_1 + \omega_3 + \omega_4 + (1+x)\omega_2, \\
p(4) &= \omega_1 - \omega_3 + \omega_4 + (1+x)\omega_2, \\
p(5) &= \omega_1 + \omega_3 - \omega_4 + (1+x)\omega_2.
\end{aligned} \tag{68}$$

Now their symmetries are more transparent under the permutation group since $S_3$ permutes $\omega_1, \omega_3$ and $\omega_4$ but leaves $\omega_2$ invariant; $p_1$ and $p_2$ are invariant under the group $S_3$ but the others are permuted to each other. Pseudo snub 24-cell consists of $N_0 = 96$ vertices, $N_3 = 24 + 24 + 96 = 144$ cells consisting of pseudo icosahedra, tetrahedra and triangular pyramids respectively, $N_1 = 432$ edges and $N_2 = 480$ faces. These numbers satisfy the Euler characteristic formula $N_0 - N_1 + N_2 - N_3 = 0$.

### 4.2 Construction of the vertices of the dual polytope of the pseudo snub-24 cell

The centers of the first three pseudo icosahedra can be taken as the weight vectors $\omega_1, \omega_3$ and $\omega_4$. The other vectors in (68) can be taken as the centers of the tetrahedron and the four pyramids up to some scale vectors. Let us denote by $c(i), i = 1, 2, ..., 5$ the centers of the respective tetrahedron and the pyramids and define

$$c(1) = \lambda p(1), \quad c(i) = \eta p(i), \quad i = 2, 3, 4, 5. \tag{69}$$

So we have eight vertices including $\omega_1, \omega_3$ and $\omega_4$. To determine the actual centers of these polyhedra the hyperplane defined by the eight vectors must be orthogonal to the vector $\Lambda$. This will determine the scale factors and the centers of the cells which can be written as:

$$\begin{aligned}
\omega_1 &= 1, \quad \omega_3 = \frac{1}{2}(1 + e_1 + e_2 - e_3), \quad \omega_4 = \frac{1}{2}(1 + e_1 + e_2 + e_3), \\
c(1) &= \frac{1+2x}{2+3x}(1+e_1), \\
c(i) &= \frac{1+2x}{3+7x+3x^2}p(i), \quad i = 2, 3, 4, 5.
\end{aligned} \tag{70}$$

These eight vectors now determine the vertices of one facet of the dual polytope of the pseudo snub 24-cell. The center of this facet is the vector $\Lambda$. This is a convex solid with 8 vertices 15(3+6+3+3) edges and 9(3+3+3) faces possessing $S_3$ symmetry.

One can generate the vertices of the dual polytope by applying the group $\left(W(D_4)/C_2\right):S_3$ on the vertices in (70) representing one of the facet of the dual polytope. One can display them as the union of three sets

$$T \cup (\frac{1+2x}{2+3x})\sqrt{2}T' \cup (\frac{1+2x}{3+7x+3x^2})R(x) \tag{71}$$

where $T$ and $T'$ are given in (5-6) and $R(x)$ can be written as follows

$R(x) =$
$\{\pm(2+x)\pm e_1\pm(1+x)e_2, \pm(1+x)\pm(2+x)e_1\pm e_2, \pm 1\pm(1+x)e_1\pm(2+x)e_2,$
$\pm(2+x)\pm e_2\pm(1+x)e_3, \pm(1+x)\pm(2+x)e_2\pm e_3, \pm 1\pm(1+x)e_2\pm(2+x)e_3,$ (72)
$\pm(2+x)\pm e_3\pm(1+x)e_1, \pm(1+x)\pm(2+x)e_3\pm e_1, \pm 1\pm(1+x)e_3\pm(2+x)e_1,$
$\pm(2+x)e_1\pm e_2\pm(1+x)e_3, \pm e_1\pm(1+x)e_2\pm(2+x)e_3, \pm(1+x)e_1\pm(2+x)e_2\pm e_3\}.$

As we have noted before for $x = -\frac{1}{2}$ in (71) what remains is the set $T$ as expected because it is the dual of the set $T'$. If $x = -\frac{2}{3}$, the dual polytope does not exist. The roots of the quadratic equation $3+7x+3x^2 =0$ are irrational numbers which are already excluded since the vertices of the pseudo snub 24-cell and its dual remain in the $D_4$ lattice for $x$ has to be rational number. One wonders whether the vertices in (72) represent any polytope familiar. If we replace $x$ by $x = \frac{-1}{y+1}$, the set $R(x \to \frac{-1}{y+1})$ in (72) takes exactly the same form of $S(y)$ apart from overall a scale factor. This implies that $R(x)$ represents another pseudo snub 24-cell. This proves that every dual of a pseudo snub 24-cell includes another pseudo snub 24-cell in addition to the 24 cells $T$ and $T'$.

The pseudo snub 24-cell, $S(x)$ turns out to be snub 24-cell whose cells are regular icosahedra and tetrahedra when $x = \tau$ or $x = \sigma$ where $\tau = \frac{1+\sqrt{5}}{2}$ and $\sigma = \frac{1-\sqrt{5}}{2}$. However $R(\tau)$ and $R(\sigma)$ represent two pseudo snub 24-cells, the mirror images of each other. In brief, when $S(x)$ represents a pseudo snub 24-cell its dual polytope involves an orbit with 96 vertices representing another pseudo snub 24-cell but if $S(x)$ is snub 24-cell in its dual the set with 96 vertices represents a pseudo snub 24-cell where either $x = -\tau$ or $x = -\sigma$ [Koca et al., 2012].

## 4.3 Another representation of the group $\left(W(D_4)/C_2\right):S_3$ and the pseudo snub 24-cell

The simple roots of the Coxeter-Dynkin diagram $D_4$ can be represented by another set of quaternions exclusive to the set of root system derived from the simple roots of Figure 11. However the new set of simple roots can be obtained from those of Figure 11 by a transformation as we will discuss later. The new set of simple roots and the weight vectors can be given as

$$\alpha_1 = \sqrt{2}e_1,\ \alpha_2 = \frac{1}{\sqrt{2}}(1-e_1-e_2-e_3),\ \alpha_3 = \sqrt{2}e_2,\ \alpha_4 = \sqrt{2}e_3,$$

$$\omega_1 = \frac{1}{\sqrt{2}}(1+e_1), \omega_2 = \sqrt{2}, \omega_3 = \frac{1}{\sqrt{2}}(1+e_2), \omega_4 = \frac{1}{\sqrt{2}}(1+e_3).$$

(73)

Note that $\frac{\alpha_i}{\sqrt{2}} \in T$, $i = 1,2,3,4$; $\frac{\omega_2}{\sqrt{2}} \in T$, $\omega_j \in T'$, $j = 1,3,4$. The generators representing the reflections with respect to the new simple roots in (73) transform the quaternionic units as follows:

$r_1 e_1 = -e_1, r_3 e_2 = -e_2, r_4 e_3 = -e_3$ leaving other quaternionic units (not shown) invariant and

$$r_2 1 = \frac{1}{2}(1+e_1+e_2+e_3) \equiv \frac{\bar{\alpha}_2}{\sqrt{2}},\ r_2 e_i = e_i \frac{\bar{\alpha}_2}{\sqrt{2}} \bar{e}_i,\ i = 1,2,3.$$

The simple roots and weights in (47-48) can be transformed to the simple roots and the weights in (73) by the transformation $a \equiv [\frac{e_2+e_3}{\sqrt{2}}, -e_2] \in [T',T]$ which transforms the quaternionic units as follows

$$1 \to \frac{1+e_1}{\sqrt{2}},$$

$$e_1 \to \frac{1-e_1}{\sqrt{2}},$$

$$e_2 \to \frac{e_2+e_3}{\sqrt{2}},$$

$$e_3 \to \frac{e_2-e_3}{\sqrt{2}}.$$

(74)

$a^2 = [1,1]$, it represents the Dynkin diagram symmetry of the Coxeter-Weyl group $F_4$ (Koca et al, 2001).

It is straightforward to show that $a[T,T]a^{-1} = [T,T]$ and $a[T,T]^*a^{-1} = [T',T']^*$. Therefore the group $\left(W(D_4)/C_2\right):S_3$ represented by $[T,T] \cup [T,T]^*$ can be transformed to the group $[T,T] \cup [T',T']^*$ in the new basis of the root system. These are the two different quaternionic representations of the same group. They are the maximal subgroups of the Coxeter-Weyl group $W(F_4)$ as we have shown in (7). The work presented in Sections 4.1-4.2 can be repeated by using the substitution (74).

The vector $\Lambda = a_2[(1+2x)+(1+x)e_1+xe_2]$ in the new choice of root system will read $\Lambda' = a_2[(2+3x)+x(e_1+e_2+e_3)]$ if an overall factor $\frac{1}{\sqrt{2}}$ is omitted.

When the group $[T, \bar{\omega}_4 \bar{T} \omega_4]$ generated by the rotation generators $r_1 r_2$ and $r_2 r_3$ are applied on the vector $\Lambda'$ one generates 12 vertices of a pseudo icosahedron which can be obtained from (61) by the substitution (24). Note that here the invariant vector under these transformations is $\omega_4 = \frac{1+e_3}{\sqrt{2}} \in T'$. Obviously $[\omega_4, \omega_4]^* \in [T',T']^*$ leaves the set of vertices of the pseudo icosahedron therefore the larger group symmetry of the pseudo icosahedron is $T_h \approx [T, \bar{\omega}_4 \bar{T} \omega_4] \cup [T', \omega_4 \bar{T}' \omega_4]^*$ which is isomorphic to the pyritohedral group. Indeed all the conjugate groups to the pyritohedral group can be written as $T_h \approx [T, \bar{s}\bar{T}s] \cup [T', s\bar{T}'s]^*, s \in T'$ implying that there are 24 pseudo icosahedra. The tetrahedral group preserving a vector $t \in T$ can be written as $T_d \approx [T, \bar{t}\bar{T}t] \cup [T', t\bar{T}'t]^*, t \in T$. For $t=1$ this group takes exactly the same form of the tetrahedral group in (20). The 96 vertices of the pseudo snub 24-cell in this bases will read

$$S'(x) = \{\pm(2+3x) \pm xe_1 \pm e_2 \pm xe_3, \pm x \pm xe_1 \pm xe_2 \pm (2+3x)e_3,$$
$$\pm x \pm xe_1 \pm (2+3x)e_2 \pm xe_3, \pm x \pm (2+3x)e_1 \pm xe_2 \pm xe_3,$$
$$\pm(1+2x) \pm (1+2x)e_1 \pm (1+2x)e_2 \pm e_3, \pm(1+2x) \pm (1+2x)e_1 \pm e_2 \pm (1+2x)e_3,$$
$$\pm(1+2x) \pm e_1 \pm (1+2x)e_2 \pm (1+2x)e_3, \pm 1 \pm (1+2x)e_1 \pm (1+2x)e_2 \pm (1+2x)e_3, \quad (75)$$
$$\pm(1+3x) \pm (1+x)e_1 \pm (1+x)e_2 \pm (1+x)e_3, \pm(1+x) \pm (1+x)e_1 \pm (1+x)e_2 \pm (1+3x)e_3,$$
$$\pm(1+x) \pm (1+x)e_1 \pm (1+3x)e_2 \pm (1+x)e_3, \pm(1+x) \pm (1+3x)e_1 \pm (1+x)e_2 \pm (1+x)e_3,$$
(even number of (+) sign)$\}$.

Note that for $x = -1, -\frac{1}{2}, 0$ this set reduces to the set $T$ as expected which represents a 24-cell. The terms with odd number of (+) sign in (75) represent the mirror image of the pseudo snub 24-cell. Therefore under the transformation in (74)

$$T \leftrightarrow T', S(x) \leftrightarrow S'(x) \text{ (even number of (+) sign)}. \quad (76)$$

## 4.4. Fibonacci sequence of the pseudo snub-24 cells

A discussion similar to that of Section 3.6 can be persuaded here. Consider the rational values of $1 \leq x \leq 2$ defined by $x_n = \frac{a_1}{a_2} = \frac{F_{n+1}}{F_n} = 1, \frac{2}{1}, \frac{3}{2}, \frac{5}{3}, \frac{8}{5}, \frac{13}{8}, \frac{21}{13}, \frac{34}{21}, \frac{55}{34}, \frac{89}{55}, \ldots$ where $F_n$ represents the general term of the Fibonacci sequence. For each value of $x_n$ we obtain a pseudo snub 24-cell in the lattice $D_4$. The coefficients of the unit quaternions in the set of vertices (59) take values from the Fibonacci sequence. For example, for $x_n = 1$ the coefficients of the unit quaternions are either $1, 2$ or $3$. Therefore the sequence of snub 24-cells will follow the sets $(1,2,3), (2,3,5), (3,5,8), \ldots, (F_n, F_{n+1}, F_{n+2})$ as the coefficients in their unit vectors. As the general term of the sequence reaches to infinity then $x_n \to \tau$ leading to the snub 24-cell. Therefore the Fibonacci sequence of the pseudo snub-24 cells approaches to the snub 24-cell whose vertices are not in the lattice $D_4$ rather in the quasilattice $H_4$.

## 5. Discussion

We have studied the extension of the pyritohedral group $\left( W(D_4)/C_2 \right) : C_2'$ in 3D to the group $\left( W(D_4)/C_2 \right) : S_3$ acting in 4D Euclidean space. Quaternionic representations of these groups setup the natural link between these two groups. We have constructed the 4D polytope with 96 vertices (pseudo snub 24-cell) and its dual polytope with 144 vertices invariant under the group. The explicit construction of the group in terms of quaternions has been worked out. The relevance of the group and the polytopes to the root lattice of the affine Coxeter group $W_a(D_4)$ has been pointed out. The fact that the group $\left( W(D_4)/C_2 \right) : S_3$ can be embedded in the 4D crystallographic group $W(F_4)$ and in the 4D quasicrystallographic Coxeter group $W(H_4)$ as maximal groups implies its transitional role between the crystallography and the quasicrystallography. We have pointed out that the dual of the pseudo snub 24-cell is the union of the sets $T$ and $T'$ representing two 24-cells dual to each other and the set $S(x)$ with 96 vertices. For rational values of $x$ the vertices belong to the lattice $D_4$. An infinite set of $S(x)$ for $1 \leq x \leq 2$ constitutes a Fibonacci sequence of pseudo snub 24-cells leading to the snub 24-cell in the limit $x_n \to \tau$. The snub 24 cell is no more in the lattice $D_4$ but belongs to the quasilattice of the Coxeter group $W(H_4)$ which can be obtained by projection of the lattice $E_8$ into 4D.